# Negative Database for Data Security

Anup Patel    Niveeta Sharma    Magdalini Eirinaki

*Computer Engineering Department*
*San Jose State University*
*anup.sjsu@gmail.com, niveeta@gmail.com, magdalini.eirinaki@sjsu.edu*

## Abstract

*Data Security is a major issue in any web-based application. There have been approaches to handle intruders in any system, however, these approaches are not fully trustable; evidently data is not totally protected. Real world databases have information that needs to be securely stored. The approach of generating negative database could help solve such problem. A Negative Database can be defined as a database that contains huge amount of data consisting of counterfeit data along with the real data. Intruders may be able to get access to such databases, but, as they try to extract information, they will retrieve data sets that would include both the actual and the negative data. In this paper we present our approach towards implementing the concept of negative database to help prevent data theft from malicious users and provide efficient data retrieval for all valid users.*

## 1. Introduction

Security related issues have been a threat to the web world since the first days of the use of web and the Internet. There are people around who have always tried to get into the internal information system and the database systems. There are organizations such as credit card companies, government agencies and security agencies that need their data secured to the highest extent. Hence such organizations want their applications to provide high security. There are numerous ways to keep data secure but, the fact is a proper answer to this question, "Do all these promised security vendors provide enough security to your data?", and this remains unanswered. There is never 100% data protection solution anywhere because of different kinds of attacks that could be vulnerable at any time and the consequences may be worst than expected, therefore no one knows the types and the severity of such attacks. In this paper we present a security framework aiming at providing data security to any application that requires high database security levels. This framework is a collection of some algorithms that manipulate the input data and store it in the database. This populated database is referred to as the *negative database*. A negative database can be defined as a database that contains huge amount of data which consists of counterfeit data along with the actual data. A few approaches that describe this concept have been proposed but have not yet been implemented to work for real world databases [5]. All the proposed algorithms had issues related to the reversibility of the data, converting back to the original form from the negative representations [3] and the fact is that they have not yet been implemented. [2, 4].

Contrary to previous work, the proposed framework makes manipulations to the original data, saves in the database and is able to retrieve them efficiently. Our main objective is the validation of a benign user and rejection of a malicious user for a particular database. There have been cases where an attack to a database can take place by writing a query in the username and password field of the login page. For a normal database, this may work and the malicious user can get access to the database of the system. On the other hand, our framework prevents a malicious user form doing so. Due to space constraints, in this paper we present the main modules and algorithms of the proposed framework. For more details the reader may refer to [6].

## 2. Architecture

In this paper we present a security framework which can be embedded at the middle layer of any web-based or stand-alone application that requires high security levels. There is no restriction on the database management system that needs to be used.

The top level architectural diagram of the implementation of the security framework concept is shown in Figure 1. A user submits a query to the database using the web browser component. The server starts processing the query and the security framework comes into action. The middle tier includes the framework that consists of several technologies combined together to give a robust safeguard to the database layer that lies below the framework. More specifically, the framework consists of four main modules, namely, *Database caching, Virtual database encryption* and *Database encryption algorithm,* along with *Negative Database conversion algorithm.*

The manipulated data is finally stored into the actual database after passing through all the components of the security framework. Thus the database contains the *negative* data, which is a manipulated form of the original one, as well as the positive data. All the queries that are used by a valid user work with the Java Hibernate technology [1]; hence the database is updated as objects and its attributes. Each table is represented as an object and the tuples are the values of the attributes of the object. Any query that is invoked towards the database will be associated with objects and its attributes; hence an invalid query will not be able to retrieve any useful information from the database. In what follows, we discuss in detail the framework's modules.

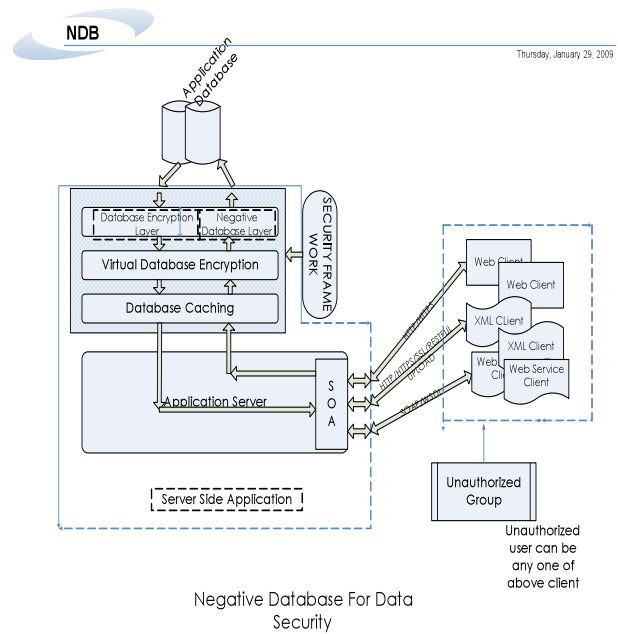

**Figure 1. System Architecture**

## 2.1 Database Caching

Database caching refers to the concept of easy access of frequently used data. It is the way in which frequently used data is stored in an easy or quick access area such as the RAM and defining specific ways of retrieving those data. The purpose of data caching has several advantages compared to the use of a simple database namely, (a) faster data access by using index to the disk and reduces disk access, and (b) higher CPU utilization as the computation time to access the data is reduced. This concept provides higher performance when there is huge amount of data access and a lot of modification to the database.

## 2.2 Virtual database encryption

Virtual database encryption refers to the use of a set of randomly generated keys attached to the objects in any hibernate scenario. It is done by generating a set of random keys using the system time and attaching it to the data in the form of objects and its attributes. This is detailed in Figure 2. The new values are then sent to the database as an attribute of object, the objects refer to the tables in the database and the value of the attributes relates to the tuples in the table.

The algorithm is invoked when a SELECT, INSERT, UPDATE or DELETE action is performed in the database, maintaining the ACID property and providing high amount of security to the data in the database along with a faster access.

```
(Call before Session is updated)
1. Read and Get data from the user
2. Get the ASCII value of the data.
   2.a.Calculate length of the data.
   2.b.Convert ASCII: charAt(i);
        Store into variable "c"
   2.c.IF c=[A-Z] and c=[a-c] and c[0-9]
       THEN (multiply by 10)
       ELSE Take variable "j" and store
            the ASCII value in j
   OR
   2.c.Concatenate all the ASCII value
       of each character with any
       primitive number "*" and store it
       in "j"
3. Get the current system date and time.
   3.a.Create an object of TimeStamp
   3.b.Get "Year", "Month", "Day"
   3.c.Get "Hour", "Minute", "Second"
   3.d.Get AM as 0 and PM as 1.
       Store into variable "z".
       IF z=0 THEN "hour=hour".
       ELSE "hour"="hour+12".
   3.e.Append all the values of "year",
       "month", "day", "hour", "minute",
       "second" and "z" and store in
       TimeStamp.
4. The TimeStamp is appended to the ASCII
value of the data which is currently stored
in variable "j" and pass the generated value
to the Database Encryption algorithm,
RSA_publicKey() layer.
```

**Figure 2. Pseudocode of Virtual database encryption**

## 2.3 Database Encryption algorithm, RSA_publicKey()

The database is always populated by data that is entered by the administrator and updated by a benign user, which could be a online banking or credit card company customer. A problematic situation may arise when a malicious user tries to update or modify the database. The example of such updates could be a SELECT query inside an INSERT query. The purpose of the encryption module along with other three modules is to provide utmost security to the data. In our framework we use the public key encryption algorithm RSA. Strong encryption makes the database more secure and reliable. The encryption algorithm is preceded by two modules which are *Database caching* and *Virtual Database Encryption algorithm*. It takes the input from the previous module and applies RSA on the data. The encrypted data is passed through the *Negative Database conversion algorithm* as shown in Figure 3 to generate encrypted multi sets of data for a single true set of data, making the database hard to query. The data encrypted in this layer is passed to the next layer called the *Negative Database Conversion layer*.

```
1.Get the input from the previous section &
pass to the RSA_publicKey()
2.Get the RSA encrypted value and pass it
to hexStringFromBytes to convert it into
hexadecimal values
  2.a.Take a variable String hex="";
  2.b.Take integer variables msb,
      lsb=0, i
  2.c.FOR EACH i ranging from
      (i=0) to (i<length of the input)
      Calculate
      msb=((int) b[i] & 0x000000FF)/16;
      lsb=((int) b[i] & 0x000000FF)%16;
      hex= hex + hexChars[msb] +
                      hexChars[lsb];
3.Pass the value generated from
hexStringFromBytes to the
getEncryptionData() method which uses MD5
  3.a.Get an instance of MD5
  3.b.Update the data using MD5 object
  3.c.Take String variable hexEncoded
      and pass the hex value using
      digest.
  3.d.Take String variable enCoded and
      pass the sub string values by
      truncating (13,15)
4. This values is passed to the next layer
called "Negative Database Conversion
Algorithm"
```

**Figure 3. Database Encryption algorithm, RSA_publicKey()**

## 3.4 Negative Database Conversion algorithm

The main concept of the framework lies on the implementation of this module. The purpose of using negative database along with the positive database is to provide extra security to the database. Any data manipulation query from the administrator or the customer is referred to as a benign user query; on the other hand any query from anyone else is always referred to as a malicious query. All the queries will go though all the modules to generate a secure database value.

The last module, also called the *Negative Database conversion algorithm*, is used to create a large set of values rather than just a single tuple. The generated sets of data are inserted in the database. Contrary to common database applications, in our negative database a malicious query will not be able to fetch the data from the database.

The term *negative* is used because of the generation of false sets of data in reference to the actual data. The actual data passes through the Virtual Database Encryption algorithm and the Database Encryption algorithm, RSA_publicKey() layer to generate the data that passes through the negative database algorithm module. The number and the type of false data generation will depend on the algorithm used.

```
1. value := Database Encryption algorithm,
RSA_publicKey()
2. key := TimeStamp from the Virtual
database encryption layer 3. The value must
be a string
4. Calculate the length of the values from
the previous layer
5. The value is either the username or the
password (the value can be any not only
username and password)
OR
5. The value could be username as value1
and password as value2
6. DO
   FOR the length of the string in step 2
   6.a.Create UserAccount Object
   6.b.Set UserName
   6.c.Similarly for all values that need
       to be changed into  negative data
   6.d.Set UserModDt(timestamp)
7. Insert <value+"*"+timestamp> tuple in
the database table
```

**Figure 4. Pseudocode of Negative Database Conversion algorithm**

After the input is manipulated using the algorithm in Figure 4, it will be saved to the database. For example, let the SSN be the attribute that we want to encrypt in the negative database. If the SSN value of customer

"Niveeta" after the Database Encryption algorithm, RSA_publicKey() layer is "4@AGD", Table 1 shows how the respective database table tuples will look like.

**Table 1. Negative Database snapshot**

| SSN | Name |
| --- | --- |
| 4 +"*"+ (Time Stamp as Key) | NIVEETA |
| @+ "*" +(Time Stamp as Key) | NIVEETA |
| A+ "*"+ (Time Stamp as Key) | NIVEETA |
| G+ "*" +(Time Stamp as Key) | NIVEETA |
| D+ "*" +(Time Stamp as Key) | NIVEETA |

The process of data retrieval from the database can be done in two ways, for a benign user and a malicious user. A benign user submits a legitimate query into the database and is able to get the correct output (e.g. <SSN, Name>). On the other hand, a malicious user writes some queries that could be vulnerable and are able to retrieve some output, but the output will comprise of numerous non-interpretable data sets, as the ones shown in Table 1.

## 3. Performance

**Database caching.** In our framework we are using system-derived timestamps as keys. Thus the complexity of the database caching algorithm O(n), when the whole database needs to be searched for a particular tuple.

**Virtual database encryption.** This layer depends on the timestamp generation and the conversion of the data into ASCII values. Thus the computation time is O(n) where n is the length of the used password.

**Negative Database conversion.** The main task of this algorithm is to generate a set of data that is to be populated to the negative database. The input from the previous module to this module is an encoded value which is an 8 digit value, hence 8 row data has to be calculated and then populated to the database. This 8 digit value remains the same length because of some constraints made in the previous module. Hence this module is not affected by the size of input, hence has O(1) complexity.

Hence the overall complexity of the security framework is O(n), when we combine the complexities of all the modules. This is however very high compared to a simple web-based application. This can be compensated by the level of security this framework provides in comparison to the low- security, high risk of data for other applications.

## 4. Conclusion

This solution of designing a security framework aims at providing high security to a web-based environment, where attacks to the database are frequent which results in theft of data and data corruption. This paper provides evidence on the retrieval of data from a *negative database*. In all the previous *Negative* Database algorithms [3], the storage of actual data as negative data was explained, however was not implemented. To the best of our knowledge, the retrieval of negative data has not yet been addressed in the literature, and this paper has made a progress on that end. In this work, we propose a framework that allows the negative representation of the original set of data, resulting in returning invalid results for malicious queries (e.g. through login). At the same time, it enables the retrieval of the original data in case of legitimate queries.

There are still many open issues related to the notion of negative databases for real world applications. This paper provides evidence on the storage and retrieval of data, for a benign user and access denial for a malicious user. This would only deal with the INSERT and SELECT query of the database. The most challenging future research based on this project could be to UPDATE the negative database. The negative database has the manipulated value as well as the timestamp of each INSERT operation. The main challenge in this scenario is to update the database value checking the timestamp; this can be done by finding the values, which is the same process as search. However the updated value should have a negative set of itself along with the system time. Building real world applications can be one of the challenging works based on the algorithms explained in this paper.